\documentclass[12pt]{article}

\usepackage{tikz}
\usetikzlibrary{arrows,petri,topaths}%
\usepackage{tkz-berge}
\usepackage{graphicx}

\usepackage{caption}
\usepackage{subcaption}

\usepackage{epsfig}
\usepackage{xfrac}
\usepackage{color}

\usepackage{graphics}
\usepackage{graphicx}
\usepackage{latexsym}
\usepackage{epsfig}

\usepackage{amsmath}
\usepackage{amssymb}
\usepackage{amsfonts}
\usepackage[lined,boxed]{algorithm2e}
\usepackage{times}
\usepackage[small,compact]{titlesec}

\def\bm{{\hbox{\bf}}}

\newtheorem{theorem}{Theorem}

\newcommand{\eat}[1]{}

\begin{document}

\title{\LARGE{On the optimality of grid cells}
\footnote{Research supported by NSF grant CCF - 1408635.}}
\author{{Christos Papadimitriou}\\
EECS, UC Berkeley}

\maketitle

\section*{Abstract}

Grid cells, discovered more than a decade ago \cite{Haft}, are neurons in the brain of mammals that fire when the animal is located near certain specific points in its familiar terrain.  Intriguingly, these points form, for a single cell, a two-dimensional triangular grid, not unlike our Figure 3.  Grid cells are widely believed to be involved in path integration, that is, the maintenance of a location state through the summation of small displacements.  We provide theoretical evidence for this assertion by showing that cells with grid-like tuning curves are indeed well adapted for the path integration task. In particular we prove that, in one dimension under Gaussian noise, the sensitivity of measuring small displacements is maximized by a population of neurons whose tuning curves are near-sinusoids --- that is to say, with peaks forming a one-dimensional grid.  We also show that effective computation of the displacement is possible through a second population of cells whose sinusoid tuning curves are in phase difference from the first.  In two dimensions, under additional assumptions it can be shown that measurement sensitivity is optimized by the product of two sinusoids, again yielding a grid-like pattern.  We discuss the connection of our results to the triangular grid pattern observed in animals.

\section{Introduction}
Grid cells \cite{Haft} are neurons in the dorsocodal medial entorhinal cortex of mammals that fire when the animal is near specific locations in its familiar environment; intriguingly, these locations form, for a single cell, a two-dimensional regular triangular grid \cite{Haft}.  Ever since their discovery, grid cells have been hypothesized to be involved in space representation \cite{Haft,GMM}, and in particular in neural algorithms {\em ``that integrate information about place, distance, and direction''} \cite{Haft}, a task usually referred to as {\em path integration} \cite{Haft,Mcnau}.  But why are neurons with grid-like tuning curves well adapted for the task of path integration?  This is the question we address in this paper.  

Path integration presumably entails the measurement of small displacements. Therefore, for path integration to be effective, measurement of small displacements has to be as accurate as possible.  What is the tuning curve, for neurons measuring small displacements, that has the highest possible sensitivity, that is, the smallest possible variance?   We show that, in one dimension, optimal measurement sensitivity is achieved through a one-dimensional grid.  

In particular, we consider a population of neurons measuring small displacements on the circle.  Working on the circle instead of the line segment, or the infinite line, simplifies the analysis by avoiding edge effects.  We assume that the tuning curves of these neurons are cyclical shifts of one another, and that the noise of the measurement is Gaussian.  We seek the tuning curve maximizing the accuracy of the measurement.  A useful surrogate of accuracy is the {\em Fisher information} \cite{Kay}, which upper-bounds the accuracy of any estimator.  We establish that the tuning curve maximizing Fisher information is a sinusoidal-like wave (see Figure 1) --- that is to say, a tuning curve whose peaks form a grid.  The frequency of the wave is that of the eigenvector corresponding to the smallest positive eigenvalue of the noise correlation matrix.  We say ``sinusoidal-like'' because, mathematically, the optimal solutions form a family of near-sinusoidal functions parametrized by any function $\psi:[0,1]\mapsto [0,1]$, with the common sinusoid corresponding to the identity function (notice the differences between the sinusoid-like waves in Figure 2).

\begin{figure}
		\centering
  		\includegraphics[width=0.3\linewidth]{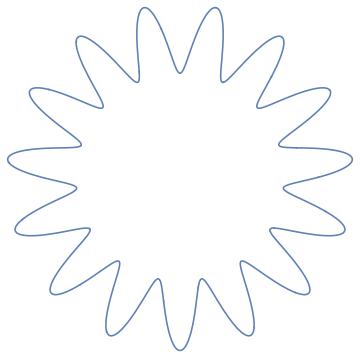}
  		\caption{The optimum tuning curve}
\end{figure}

But how can the displacement be read out from the change in spiking rates in such a population?  We notice that the actual displacement can be computed with the help of a second cell population which is identical to the first, albeit with tuning curves phase-shifted by $90^{\rm o}$.   

This last observation about the computation of displacement is relevant when exploring how our analysis can be extended to two dimensions.   Again, in order to avoid edge effects we are working on the torus (the unit square with its opposite sides identified).  Under quite hefty assumptions (of independence of both correlation and tuning curves, as well as of independent optimization of sensitivity in each direction) it can be shown that the optimum tuning curve in two dimensions is indeed a grid.  We speculate that the triangular grid may be the result of optimization of measurement sensitivity under an additional constraint, and that constraint may be computational:  In two dimensions, the problem of inferring the displacement from tuning curve change requires not one additional phase-shifted population of cells, but {\em three} such populations.  The interface of the four phase-shifted populations is best achieved through a triangular grid (Figure 3).  

\paragraph{Related work:}
Over the past decade, there has been much theoretical investigation of grid cells, their origin, and their role.  It has been noted that a grid-like pattern can result from the interference of two, or possibly three, sinusoidal waves \cite{Burg}, while periodic tuning curves on the circle \cite{N} and the torus \cite{Mcnau} can be generated by neural networks (see Figure 3A and B in \cite{GMM}); however, these models were not proposed in the context of optimizing the accuracy of measurement.  It has also been shown that continuous attractor models can generate triangular grid-like responses \cite{BF}, while experimental data are consistent with a 2-dimensional response of the population \cite{YBB}.  Grid cells were interpreted in \cite{SF} and, in a different way, in \cite{FBB}, as very efficient novel neural codes for encoding position and velocity, and in \cite{MHS} it is shown that in this arena grid cells are more apt than place cells; these works are methodologically close to ours in that they also employ Fisher information for their analysis and comparisons --- without, however, seeking the tuning curve design that maximizes it.  Cells with one-dimensional grid-like firing patterns have also very recently come up in the analysis of the responses of animal grid cells to one-dimensional environments \cite{F}, characterized as projections (``slices'') of a two-dimensional lattice to the one-dimensional circumstances of the experiment.  Finally, recently it was claimed in \cite{M} that the tuning curve in one dimension with maximum Fisher information is a sinusoid curve, while the product of two such curves is optimum in two dimensions, results very similar to ours; unfortunately (as pointed out in Sections 2 and 3) the mathematical development in that manuscript contains significant gaps. 



\section{The optimal tuning curve is periodic}
\label{sec:framework}
\label{subsec:framework-outline}
Consider a population of $N$ neurons measuring a small angular displacement at a point on the circle.   We assume that the tuning curves of the $N$ neurons are identical, albeit shifted by multiples of the angle $2\pi\over N$.  We seek the tuning curve that maximizes the sensitivity of measuring small displacements. 

The neurons respond to an angular stimulus $\theta\in [0,2\pi]$, and the tuning curve of the $i$th neuron is denoted by $f_i(\theta)$; we assume that the tuning curves are identical but shifted by multiples of ${2\pi\over N}$, that is, $f_{i+1\bmod N}(\theta) = f_i(\theta + {2\pi\over n})$. 
The average population activity caused by a stimulus is thus the vector $\bm{f}(\theta)=(f_1(\theta), f_2(\theta),...,f_N(\theta))$.   The derivative $d{f}_i\over d\theta$ is denoted $\dot f_i$, and we denote by $\bm{\dot f}(\theta)$ the vector $(\dot f_1(\theta), \dot f_2(\theta),...,\dot f_N(\theta))$.

A stimulus $\theta$ results in the response $r_i(\theta)=f_i(\theta)+\eta_i$, for $i=1,\ldots,N$, where $\eta_i$ is Gaussian noise.  The values of noise at different neurons, $\eta_i$ and $\eta_j$, are correlated, and this correlation is assumed to be independent of $\theta$, and denoted  $C_{i,j}$, a quantity that depends on the distance on the ring of the neurons $i$ and $j$. It follows that the noise correlation matrix $C$ is both circulant and symmetric.

Importantly, we also assume that the total signal power ---  the sum of the squares of the slopes of the tuning curves --- is bounded from above by a constant, which we take to be one.  Thus $\dot{\bm{f}} (\theta)^T \dot{\bm{f}} (\theta)\leq 1$ for all $\theta$.

If the stimulus changes from $\theta$ to $\theta + \Delta \theta$, this results in population activity $\bm{r}(\theta+\Delta \theta)$.    The goal of the decoding system is to estimate $\Delta \theta$ from the change in the population response --- that is, from $\bm{r}(\theta) - \bm{r}(\theta + \Delta \theta)$.   To do this as effectively as possible, the overall variance of the measurement must be as small as possible.  Instead of this ``overall variance'', it is convenient in this context to work with the  {\em Fisher information} of the population, a function of the tuning curves and the correlation matrix which is known, by the Cramer-Rao theorem \cite{Kay}, to bound from above the accuracy of any any unbiased estimator.   We seek the tuning curves $f_i(\theta)$ with the largest {Fisher information} under correlation matrix $C$.

Under Gaussian noise, and the assumption that $C$ is nonsingular, it is well known \cite{Kay,YS} that the Fisher information can be written as follows:

$$\bm{I}(\theta) = {\bm{\dot f}} (\theta)^{T} \bm{C}^{-1} {\bm{\dot f}} (\theta)\eqno{(1)}$$

Thus, we seek the vector $\bm{\dot f}$ satisfying $\bm{\dot f}^T\bm{\dot f}\leq 1$ that maximizes the right-hand side of (1).  

Furthermore, $\bm{C^{-1}}$ is also symmetric, and its eigenvalues are the inverses of the eigenvalues of $\bm{C}$, while its eigenvectors are the same as those of $\bm{C}$.  

Recall now the Courant-Fischer theorem \cite{Mey} (stated below for the case of real symmetric matrices and the largest eigenvalue only):

\begin{theorem} {\bf (Courant-Fischer, 1953)} If $A$ is symmetric, then the vector $x$ in the unit ball $x^Tx\leq 1$ that maximizes $x^TAx$ is the eigenvector corresponding to the largest eigenvalue of $A$.  
\end{theorem}

Comparing with equation (1), we conclude that the optimum tuning curve vector $\bm{f}$ has derivative $\bm{\dot f}$ equal to the eigenvector corresponding to the {\em smallest} positive eigenvalue of $\bm{C}$ (the inverse of the largest eigenvalue of $C^{-1}$).   What is this eigenvector?

Since $\bm{C}$ is circulant and symmetric, it is well known  \cite{Mey} that each eigenvalue $\lambda_k$, for $k=0,\ldots, {N\over 2}-1$, has multiplicity two, and the two corresponding eigenvectors are the two sinusoidal waves $v_k$ and $w_k$:

$$v_k=[\cos(0),\cos(k\delta),\cos(2k\delta),\ldots, \cos((N-1)k\delta)] \hbox{\rm \ \  and}$$
$$w_{k}=[\sin(0),\sin(k\delta),\sin(2k\delta),\ldots, \sin((N-1)k\delta)],$$
where $\delta={2\pi\over N}$ and $k=0,1,\ldots,{N\over 2}-1$.

We conclude that the optimum tuning curve vector has derivative of the form
$$\bm{\dot f}(\theta)=\alpha_k(\theta)v_k + \beta_{k}(\theta)w_{k},$$

where $\alpha_k^2(\theta) + \beta^2_{k}=1$ and the smallest positive eigenvalue of $C$ is $\lambda_k$.  We now apply the change of variables \footnote{The otherwise similar argument in \cite{M} does not contain this step, and as a result it is incomplete and the full spectrum of optimal solutions (see Figure 2) is missed.} 
$\alpha_k(\theta) =\sin(\phi(\theta)), \beta_{k}(\theta) =\cos(\phi(\theta))$ to obtain

$$\dot{\bm{f}}_i(\theta)=\sin(\phi(\theta))\cos(ik\delta)+\cos(\phi(\theta))\sin(ik\delta) =\ 
\sin(\phi(\theta)+ik\delta).\eqno{(2)}$$

Recall that we are assuming that the tuning curves of the $N$ neurons are identical, albeit shifted by $\delta={2\pi\over N}$; that is, for all $i$ and $\theta$, $f_{i+1}(\theta) = f_i(\theta+\delta)$, and thus
$\dot{f}_{i+1}(\theta) = \dot{f}_i(\theta+\delta)$.  Substituting into (2) we conclude that $\sin(\phi(\theta)+(i+1)k\delta) = \sin(\phi(\theta+\delta)+ik\delta)$
for all $\theta$, or equivalently 
$$\phi(\theta+\delta) = \phi(\theta) +k\delta + 2n\pi, \eqno{(3)}$$  
for some integer $n$.  The simplest solution of (3) is $\phi(\theta)=(k+nN)\theta +c$ for all $\theta$ and for some constant $c$ (which we take zero without loss of generality) and integer $n$.  It follows that ${\dot f}_i(\theta)= \sin(K\theta+ik\delta),$ and thus
$${f}_i(\theta)= {1\over K}\cos(K\theta+ik\delta)+\left|{1\over K}\right|,\eqno{(4)}$$
where $K=k+nN$ for some integer $n$ (positive, zero, or negative), and we took the constant of the integration to be $|{1\over K}|$ so the tuning curve takes only positive values.

However, (3) has many more solutions.  Let $\psi(\theta)$ be any function mapping $[0,\delta]$ to the reals, and, for any $\theta\in[0,2\pi]$, define $\theta'\in [0,\delta]$ and $\theta''\in \{0,\delta, 2\delta,\ldots 2\pi-\delta\}$ by the equation $\theta = \theta'+\theta''$.  Then the function 
$$\dot f(\theta) = \sin(\psi(\theta')+ \theta'')$$
is also the derivative of an optimal tuning curve.  Indicatively, in Figure 2 we show the tuning curves resulting from six simple functions $\psi$, including $\psi(x)=x$ (the true sinusoid).  

\begin{figure}[ht] 
 \hfil
  \begin{subfigure}[b]{0.25\linewidth}
    \centering
  		\includegraphics[width=\linewidth]{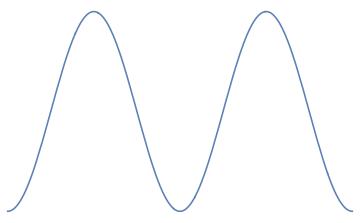}
  		\caption{$\psi(x) = x$}
    \label{fig7:a} 
    \vspace{4ex}
  \end{subfigure}
  \hfil
  \begin{subfigure}[b]{0.25\linewidth}
    \centering
  		\includegraphics[width=\linewidth]{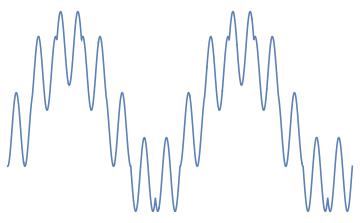}
  		\caption{$\psi(x) = 10x$}  
    \label{fig7:b} 
    \vspace{4ex}
  \end{subfigure}
  \hfil
 \begin{subfigure}[b]{0.25\linewidth}
    \centering
  		\includegraphics[width=\linewidth]{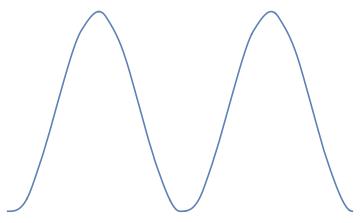}
  		\caption{$\psi(x) = \frac{x^2}{\delta}$}  
    \label{fig7:c} 
    \vspace{4ex}
  \end{subfigure}
  \hfil
  \vskip\baselineskip 
  \begin{subfigure}[b]{.25 \linewidth}
    \centering
  		\includegraphics[width=\linewidth]{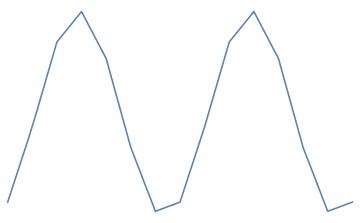}
  		\caption{$\psi(x) = 1$}
    \label{fig7:d} 
  \end{subfigure}
  \hfill
    \begin{subfigure}[b]{0.25\linewidth}
    \centering
  		\includegraphics[width=\linewidth]{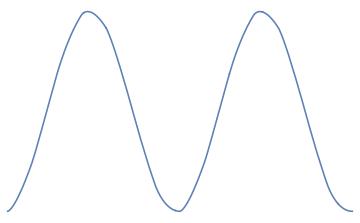}
  		\caption{$\psi(x) = \sqrt{x \delta}$}
    \label{fig7:e} 
  \end{subfigure}
  \hfill
  \begin{subfigure}[b]{0.25\linewidth}
    \centering
  		\includegraphics[width=\linewidth]{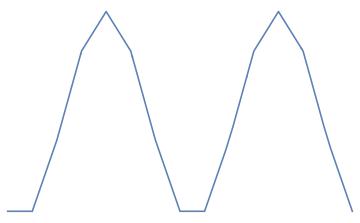}
  		\caption{$\psi(x) = 0$}
    \label{fig7:f} 
  \end{subfigure} 
  \caption{Six optimum tuning curves}
  \label{fig7} 
\end{figure}

\section{Two Dimensions}
In view of the one-dimensional result, one may suspect that grid-like structures may also be optimal in two dimensions.  Intuitively, it is tempting to try and reduce the two-dimensional case to the one-dimensional case just solved, and show that the optimum two-dimensional tuning curve must be the product of two sinusoids, and therefore a grid.  Unfortunately, this matter turns out to be quite a bit more complicated. 

Assume a population of neurons measuring displacements on the torus (the product of two circles, or, equivalently, a square with opposite edges identified in the parallel way), whose tuning curves are shifts of one another along some lattice on the torus defined by the unit vectors along two different directions $x$ and $y$ (not necessarily orthogonal).  We make some additional assumptions:

\begin{itemize}
\item The noise correlations in the $x$ and $y$ directions have the same form and are independent of each other.  That is, the correlation tensor decomposes into the product of two identical correlation matrices.

\item We further assume that the tuning curve of the neurons can also be decomposed as the product of identical one-dimensional tuning curves in the $x$ and $y$ dimensions.  It then follows that the Fisher information in the $x$ dimension has the form
$${I_x}(x) = \left[\dot{f_x}(x)^T C^{-1}\dot{f_x}(x)\right]\cdot \left[(f_y(y)^T C^{-1} f_y(y))\right],\eqno{(5)}$$
and similarly for the $y$ direction.

\item Identifying the optimum tuning curve, even under these assumptions, is still ill-defined, because of the possibly unbounded second factor in (5): $f_y$ must be obtained by integrating $\dot{f_y}$, a step that introduces an unbounded integration constant.  We could of course impose an upper bound on $f_y$ --- a justified assumption since neurons cannot fire at arbitrarily high rates, --- but then the optimization problem becomes an intractable one, involving integral inequality constraints\footnote{This difficulty is ignored in \cite{M}.}.  We can obtain a meaningful solution only under one additional assumption:  That the Fisher information in each of the $x$ and $y$ directions is maximized {\em independently of the other direction.}  That is, the overall sensitivity is not maximized, and instead the sensitivities along the two directions $x$ and $y$ are maximized separately, yielding an overall suboptimal solution.  This is not 
implausible, if one considers the independent evolution of two separate modules, each measuring displacement in one of the two directions. 
\end{itemize} 

Under these assumptions, the result for the one-dimensional case does generalize immediately, and the tuning curve maximizing the Fisher information at all stimuli turns out to be the outer product of near-sinusoidal waves in the $x$ and $y$ directions.   If the two directions form an angle of $120^{{\rm o}}$, the familiar triangular grid results.
The idea that the triangular firing field structure can result from the interference of two oscillations has been suggested before \cite{Burg}; however, the advantage of this structure was unclear.

But why should the two directions $x$ and $y$ be at an angle of $120^{{\rm o}}$ to form the familiar triangular grid observed in \cite{Haft}?  One possible answer comes from algorithmic considerations,  discussed next. 

\section{Computing the displacement}
Consider a population of cells around the circle as in the previous section with the sinusoid tuning curve $f(\theta)$ in (4) above, measuring (under noise) the change in firing rate $\Delta f = f(\theta)-f(\theta+\Delta\theta)$.
What is the mechanism whereby the displacement  $\Delta\theta$ is inferred from the measurement of $\Delta f$?  This seems problematic, since the value of the stimulus $\theta$ appears to be needed, and keeping track of $\theta$ is the purpose of path integration... But upon closer consideration, we note that the decoding mechanism does not quite need the stimulus $\theta$, but just its cosine.   Recall that, as $\Delta\theta$ goes to zero, $\Delta f = \dot f \Delta \theta$, where $f = \sin(K\theta + \hbox{\rm const})$.  Hence,
$$\Delta\theta = \Delta f{1\over K\cos(K\theta)}.\eqno{(6)}$$  
In the one-dimensional case, this additional information can be obtained with a very simple architecture:  Suppose that there is a {\em second} population of neurons, with identical tuning curve $g(\theta)$ with the primary population, albeit shifted by an angle $\alpha\neq 0, \pi$; say $\alpha = 90^{{\rm o}}$.  Then this new population yields a similar equation, with $\sin$ replacing $\cos$ because of the phase shift:
$$\Delta\theta = \Delta g{1\over K\sin(K\theta)}.\eqno{(7)}$$  
Thus we have two trigonometric equations for the two unknowns $\Delta\theta$ and $\cos(K\theta)$, which can be easily solved:  By dividing (6) by (7) we note that $\tan(\theta)={\Delta g\over \Delta f}$, and hence
$$\Delta\theta = \Delta f{1+({\Delta g\over \Delta f})^2\over K}.$$
That is, the displacement measurement can be computed from the two populations.  We conclude that, in one dimension, a second identical population of neurons shifted by $90^{{\rm o}}$ suffices for an effective readout.

Now in two dimensions, the equivalent of (6) is
$$\Delta \theta_x K \cos (K\theta_x) + \Delta \theta_y K \cos (K\theta_y)
= \Delta f.$$
Notice that now there are {\em four} unknowns ($\Delta \theta_x$,  $\cos (K\theta_x)$, $\Delta \theta_y$, and $\cos (K\theta_y)$).  We conclude that {\em three} additional populations of neurons seem to be required, with different two-dimensional shifts from the original one.  The most natural way to implement such a scheme is through shifts in three directions, forming equal angles of $120^{{\rm o}}$ with each other (see Figure 3).  Hence the familiar triangular grid may be the most natural way to implement this mechanism.  Further analytical articulation of this point is the subject of future work.  

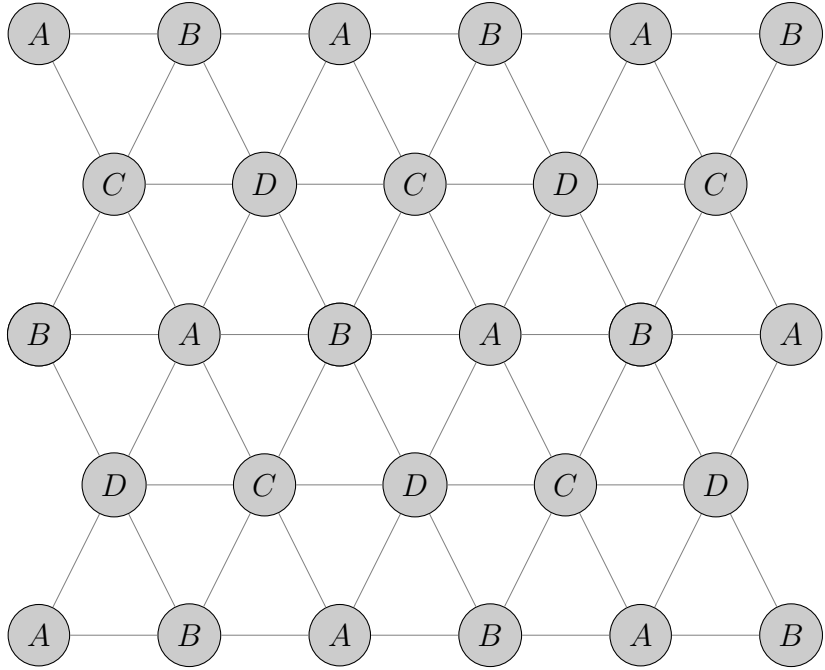
\begin{figure}
\centering
\begin{tikzpicture}[darkstyle/.style={circle,draw,fill=gray!40,minimum size=20}]
  \foreach \x in {0,2,4}
    \foreach \y in {0,4}
       	\node [darkstyle]  (\x\y) at (2*\x,2*\y) {$A$};
       	
  \foreach \x in {1,3,5}
    \foreach \y in {0,4}
       	\node [darkstyle]  (\x\y) at (2*\x,2*\y) {$B$};
       	
  \foreach \x in {0,2,4}
    \foreach \y in {2}
       	\node [darkstyle]  (\x\y) at (2*\x,2*\y) {$B$};
       	
  \foreach \x in {1,3,5}
    \foreach \y in {2}
       	\node [darkstyle]  (\x\y) at (2*\x,2*\y) {$A$};
       	
  \foreach \x in {0,2,4} 
    \foreach \y in {2}
       	\node [darkstyle]  (\x\y) at (2*\x,2*\y) {$B$};
       	
  \foreach \x in {0,2,4}
    \foreach \y in {1}
       	\node [darkstyle]  (\x\y) at (2*\x+1,2*\y) {$D$};
  
  \foreach \x in {1,3}
    \foreach \y in {1}
       	\node [darkstyle]  (\x\y) at (2*\x+1,2*\y) {$C$};
  
 \foreach \x in {0,2,4}
    \foreach \y in {3}
       	\node [darkstyle]  (\x\y) at (2*\x+1,2*\y) {$C$};
  
  \foreach \x in {1,3}
    \foreach \y in {3}
       	\node [darkstyle]  (\x\y) at (2*\x+1,2*\y) {$D$};
  
   \foreach \x in {0,...,4}{
     	\pgfmathtruncatemacro{\right}{\x+1}
     	\draw[gray,very thin] (\x4)--(\right4);}
       	
   \foreach \x in {0,...,4}
     \foreach \y in {0,2}{
     	\pgfmathtruncatemacro{\right}{\x+1}
     	\pgfmathtruncatemacro{\up}{\y+1}
     	\draw[gray,very thin] (\x\y)--(\right\y)  (\x\y)--(\x\up);}
     	
   \foreach \x in {0,...,3}
     \foreach \y in {1,3}{
     	\pgfmathtruncatemacro{\right}{\x+1}
     	\pgfmathtruncatemacro{\up}{\y+1}
     	\draw[gray,very thin] (\x\y)--(\right\y)  (\x\y)--(\x\up);}
     	
   \foreach \x in {0,...,4}
     \foreach \y in {1,3}{
     	\pgfmathtruncatemacro{\down}{\y-1}
     	\pgfmathtruncatemacro{\up}{\y+1}
     	\pgfmathtruncatemacro{\right}{\x+1}
     	\draw[gray,very thin] (\x\y)--(\right\down)  (\x\y)--(\right\up);}
  
  \draw[gray,very thin] (42)--(41) (44)--(43);
\end{tikzpicture}

\caption{Four interlaced populations of grid cells}
\end{figure}

\section{Discussion}
What is the origin and utility of the grid cells' distinctive firing field, and what does it have to do with path integration?  We have shown that, in one dimension, the tuning curve that optimizes the accuracy of displacement measurements is a near-sinusoid wave, whose peaks naturally form a one-dimensional grid.  In two dimensions, we needed several further assumptions in order to show that the optimum tuning curve is the product of two sinusoidal waves in two non-parallel directions.  If these two directions form an angle of $120^{\hbox{{\small o}}}$, the familiar triangular grid results.  We have also presented ideas about a possible mechanism for computing the change in position from the change in the response.  For two dimensions, our proposed mechanism predicts the existence of four populations of grid cells with regularly displaced firing fields, and suggests that the triangular architecture may be the optimum solution of the joint problem of maximizing both sensitivity and accuracy of decoding.

One may further speculate about grid cells for {\em three dimensions} (relevant for animals such as bats and sea mammals).  Here, our analysis predicts a total of {\em six} populations, and these must be interlaced in a way analogous to that in Figure 3.  The arrangement in Figure 3 works because the simple triangular lattice of one population, say population A, if copied three times and appropriately shifted, has the property that each point has at least one neighbor from each of the other three populations.   This suggests the following technical question in 3-dimensional geometry:  Is there a lattice in three dimensions with the property that, if it is duplicated six times, each point of each copy of the lattice is adjacent to at least one point from each of the other five copies?  It turns out that the lattice generated by the vectors $(0,3,0), (0,0,2), (1,1,1)$ has this property: Notice that the corresponding matrix has determinant $6$, suggesting that six copies of the lattice can be arranged in space, and it is easy to check that each point of each copy is at distance one from a point from each of the other copies.  Therefore, intriguing  compromises do exist between the design of a neural architecture for path integration in three dimensions and the realities of three-dimensional geometry.

\paragraph{Acknowledgment:}  Many thanks to Reza Moazzezi for fruitful collaboration during the early stages of this work, to Umesh Vazirani for illuminating discussions, and to Christos-Alexandros Psomas for help with the figures.

\end{document}